\documentclass[conference,a4paper]{IEEEtran}
\IEEEoverridecommandlockouts
\usepackage{cite}
\usepackage[hyphens]{url}
\usepackage[font=small,labelfont=bf]{caption} 
\usepackage{amsmath,amssymb,amsfonts}
\usepackage{algorithmic}
\usepackage{graphicx}
\usepackage{textcomp}

\usepackage{xcolor}
\usepackage{multirow}
\usepackage{subfigure}
\def\BibTeX{{\rm B\kern-.05em{\sc i\kern-.025em b}\kern-.08em
    T\kern-.1667em\lower.7ex\hbox{E}\kern-.125emX}}

\newcommand{\ie}{\emph{i.e.}, }
\newcommand{\eg}{\emph{e.g.}, }

\usepackage{enumitem}
\usepackage{verbatim}
\usepackage{tikz}

\usetikzlibrary{shapes,arrows,positioning,patterns}

\IEEEoverridecommandlockouts
\IEEEpubid{\makebox[\columnwidth]{978-1-7281-5684-2/20/\$31.00~\copyright{}2020 IEEE \hfill} \hspace{\columnsep}\makebox[\columnwidth]{ }}

\begin{document}

\title{How Crisp is the Crease? A Subjective Study on Web Browsing Perception of Above-The-Fold}

\author{\IEEEauthorblockN{ Hamed Z. Jahromi}
\IEEEauthorblockA{\textit{School of Computer Science} \\
\textit{University College Dublin}, Ireland \\
hamed.jahromi@ucdconnect.ie}
\and
 \IEEEauthorblockN{Declan T. Delaney}
 \IEEEauthorblockA{\textit{School of Electrical and Electronic Engineering} \\
\textit{University College Dublin}, Ireland \\
 declan.delaney@ucd.ie}
 \and
 \IEEEauthorblockN{Andrew Hines}
\IEEEauthorblockA{\textit{School of Computer Science} \\
\textit{University College Dublin}, Ireland \\
andrew.hines@ucd.ie}
 }

\maketitle

\begin{abstract}
Quality of Experience (QoE) for various types of websites has gained significant attention in recent years. In order to design and evaluate websites, a metric that can estimate a user's experienced quality robustly for diverse content is necessary. SpeedIndex (SI) has been widely adopted to estimate perceived web page loading progress. It measures the speed of rendering pixels for the webpage that is visible in the browser window. This is termed Above-The-Fold (ATF). The influence of animated content on the perception of ATF has been less comprehensively explored. In this paper, we present an experimental design and methodology to measure ATF perception for websites with and without animated elements for various page content categories. We found that pages with animated elements caused people to have more varied perceptions of ATF under different network conditions. Animated content also impacts the page load estimation accuracy  of SI for websites. We discuss how the difference in the perception of ATF will impact the QoE management of web applications. We explain the necessity of revisiting the visual assessment of ATF to include the animated contents and improve the robustness of metrics like SI. 
\end{abstract}

\begin{IEEEkeywords}
Visual Progress, Perceived Progress, Web QoE, Network Impairments
\end{IEEEkeywords}

\section{Introduction}
\label{introduction}
% ------------ RQs --------------
% (1) to establish a range of annotated visual completion  and above-the-fold (VC ) times. 
%(2) How the network factor impacts the perception of ATF 
%(3) to evaluate the level of  inter-annotator agreement  on the visual completion point. 
%-------------------------------
% \todo{Define the definitions and terms properly. Make sure it is consistent throught the paper}\\

% \todo{Introduction to web QoE and QoE management for the web}

 Web-based applications have always been of great interest to service  providers and facilitated the rise of cloud computing~\cite{miller2008cloud}. Web-based applications allow businesses and users to utilise software through the cloud and become less dependent on individual users' operating system and computing power. The success of such web-based applications, however, is partly dependant on how the users perceive quality (Quality of Experience). The concept of Quality of Experience~(QoE) refers to ``the degree of delight or annoyance of the user of an application or service'' \cite{qualinet2012qoe}.  QoE considers multiple Influential Factors (IFs): context, user, content and system. As a result, QoE provides, insights into user quality perception and satisfaction~\cite{barakovic2017survey,bocchi2016measuring}. 
 Measuring the performance of web browsing is essential to understand how to improve user experience and satisfaction. Researchers have been actively developing metrics and models to estimate QoE of web browsing (Web QoE). QoE metrics and models help application and service providers to identify the potential problem areas and to improve QoE of the provided services (QoE Management). QoE models estimate the perceived quality using mapping functions between application Key Performance Indicator (KPI) metrics and the subjective QoE experimental results (\eg Mean Opinion Score). Waiting time metrics, both time instant and time integral, are valuable KPIs for web QoE estimations. Analysis has shown that the less time a user waits, the better their rating of perceived quality~\cite{egger2012time}. Time instant metrics are computed based on measuring the time instant of an event occurred during web application loading process (i.e. Time To First Byte (TTFB), Time to Data Object Model (DOM) Load  and Page Load Time (PLT)). However they do not fully capture a user's flow of experiences for a series of events. Interestingly, time integral metrics are better aligned to QoE as they estimate the application quality, covering the entire loading process of web browsing~\cite{da2018narrowing}. For instance, Speed Index (SI) is used to estimate how fast a web page is visually completed, starting  from the time that a URL is requested until the time that the visual content is completely painted on the current viewport of the screen. The time that the web page is visually completed is called the Above-The-Fold (ATF) time~\cite{brutlag2011above}. The term was coined from newspaper broadsheets, where only the top half of the front page (above the fold) is visible on the newsstand. Determining ATF time is an important factor influencing the estimation of time integral metrics~\cite{saverimoutou2018web} and QoE estimation models ~\cite{hobetafeld2018speed}. Time integral metrics in general do not consider the time it takes to load the contents behind the scene (\eg JavaScript libraries, CSS or below the fold contents). The page can be visually completed while below the fold content is still loading (\ie PLT occurs after ATF). A difference in ATF time and PLT can be caused due to numerous reasons such as: limited network capacity, application content type/size and available computing resources. 

The web has evolved to serve dynamic, multimedia and immersive content~\cite{martinez2016evolution}. For a web application with an animation in the ATF content area, the computed ATF time can be prolonged until the animation stops or the page load event occurs (PLT)~\cite{speedindex}. This extended ATF time can potentially impact the quality estimation of time integral metrics and QoE models. It is important to identify the point in the page loading time process that users perceive the ATF as complete. It is also important to understand whether user perception of ATF changes if the web page takes longer to load due to network service factors. To the best of our knowledge, there is no systematic study/dataset that explores how users perceive ATF when they face a longer waiting time caused by network or content factors (i.e. animated and static only elements.)

To address this gap, we performed a user study to assess how a longer loading time caused by a network factor (bandwidth) impacts the perception of ATF, from the perspective of time and visual progress measurements. We conducted a subjective study where participants watched recorded videos of loading for seven top visited national commercial websites based on three different bandwidth conditions. For each video, the subjects selected the video frame at which they believed ATF was visually completed.
The four key contributions of this paper are: 1) A dataset annotated for ATF comprising website page load videos for a range of visual completion time and progress based on content and network bandwidths.
2) Demonstrating the influence of the network as a factor on the perception of ATF. 3) Exploring the influence of animated content on the perception of ATF. 4) Discussing the impact of estimated ATF on web performance metrics and QoE models. 
%The remainder of this paper is organised as follows.
%In Section~\ref{relatedwork}, we present background information and related work on visual metrics in the context of Web QoE. Then, we describe the methodology and design of the conducted study in Section~\ref{methodology}. Next, we present and discuss the results of the user study in Sections~\ref{results} and \ref{sec:discussion} respectively. Finally, we conclude our study and identify future work in Section~\ref{conclusion}.

\section{Background and Related Work}
\label{relatedwork}
% ------- BACKGOUND ----------
\begin{figure}[!tp]
\centering
\includegraphics[clip,width=\columnwidth]{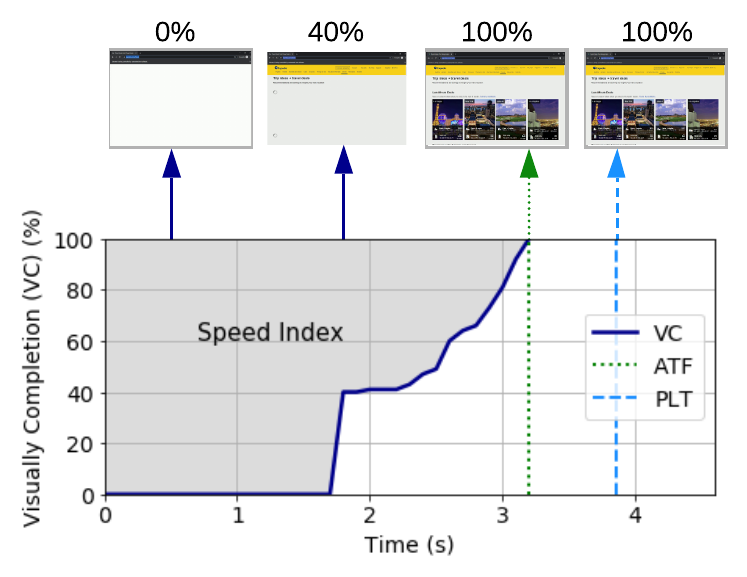}

%\captionsetup{belowskip=0pt}
\caption{Illustration of the Visually Complete (VC) time (x-axis) and progress (y-axis) with example page rendering. SpeedIndex calculates the area above the curve from page request time, $t=0$, until the time at which the ATF is 100\% VC (i.e. ATF time). VC progress usually occurs before the Page Load Time (PLT) event.  
}
\label{fig:def:siatf}
\vspace*{-5mm}
\end{figure}
Performance of web applications are measured using quality metrics divided into two distinct categories: 
\begin{enumerate}
    \item \textbf{Time instant metrics:} Computed based on measuring the time instant that an event occurred during web page loading process. For example, Time to First Byte, Time to DOM Load, Time to First Paint, Largest Content-full Paint, ATF time, Time to Last Paint, Page Load Time (PLT)~\cite{jahromi2020}.
    \item \textbf{Time integral metrics:} Used to quantify how fast a web page loads by integrating all events of a given type, tracked during the progress of a web page (using mathematical integration). For instance, SpeedIndex~\cite{speedindex}, ObjectIndex and ByteIndex~\cite{bocchi2016measuring}. 
\end{enumerate}

One criticism across much of the literature on the use of time instant metrics is that while they measure the exact time at which an event occurred, the user experience is more relevant to the flow experience than an occurrence of a particular event during the web page loading process. The shortcoming of time instant metrics motivated  researchers to explore the effectiveness of time integral and visual metrics. Interestingly, ATF time is a time instant metric~\cite{subramanian2014techniques}  that has been widely used in as an input to time integral metrics such as SpeedIndex (SI)~\cite{speedindex}, Perceptual SpeedIndex ~\cite{gao2017perceived}, ObjectIndex and ByteIndex~\cite{da2018narrowing}.  

\subsection{Visual Quality Metrics}
SI was introduced by Google in 2012~\cite{speedindex}. SI is a speed score (expressed in milliseconds) that estimates how fast the ATF content of a web page is visually painted (see Fig.~\ref{fig:def:siatf})~\cite{bocchi2016measuring}. A lower SI score points to a better perceived performance. SI uses the following equation to estimate the loading speed: 

\begin{equation}
X=\int_{0}^{t_\textrm{end}} (1 - x(t))dt
\end{equation}

where  $X$ is the estimated speed, $t_{end}$ is the time the  last event occurs, and $x(t) \in [0, 1]$ is the time evolution of the Visual Completion (VC) to reach $t_{end}$ (as shown in Fig.~\ref{fig:def:siatf}, it calculates area above the curve). For example, \textit{PLT} is generally considered as the $t_{end}$ time while $x(t)$ is the VC progress ratio of the web page over time. The VC progress ratio of SI is calculated based on a comparison of Mean Pixel Histogram Difference (MPHD) between the current state of the web page at time $t$ and the state of the page at the PLT time.  SI uses a series of snapshots (at a rate of 10 frames per second) from the time that the URL is requested until PLT. The frames are analysed in the same order to determine the VC progress ratio over time. The time that VC progress reaches to 100\% is referred to as ATF time~\cite{brutlag2011above}. 

VC time and progress are related factors influencing the result of SI. If the time increases and the VC progress has not reached to 100\% (\ie ATF has not completed the page paint), the area above the curve also increases.

Saverimoutou et al. have proposed TFVR (Time for Full Visual Rendering), a browser-based  method to determine ATF time~\cite{saverimoutou2018web}. TFVR calculates the ATF time based on the loading time of the visible portion at first load of a web page. TFVR computes ATF by extracting loading events and rendering timing from the browser HTTP Archive (HAR) file.  

Da Hora et al. argue that it is computationally expensive to measure  ATF time using MPHD analysis~\cite{da2018practical}. They proposed a metric called Aproximated Above-The-Fold (AATF) time. Similar to TFVR, AATF estimates ATF time from the browser's heuristics without requiring image processing.

Asrese et al. used a thresholding technique to determine ATF time over cellular networks~\cite{asrese2019measuring}. They argue that using a MPHD analysis, a three seconds threshold (no visual change in three seconds) is sufficient to determine ATF time for a wider range of websites with different content features. 

\subsection{Visual Metrics and Web QoE Models}
Researchers utilise different variations of ATF and SI metrics for Web QoE analysis. SI and ATF have been used as objective quality metrics in QoE models to estimate the perceived quality of the end users~\cite{bocchi2016measuring,hobetafeld2018speed}. The perceived quality is commonly  measured using subjective user ratings, e.g. 5-point or 7-point Absolute Category Rating (ACR) scales~\cite{albaum1997likert}. A Mean Opinion Score (MOS) is derived from an arithmetic mean of the users' rating for a given condition. 

Fiedler et al. proposed IQX model which refers to exponential Interdependency of Quality of eXperience and QoS (IQX)~\cite{fiedler2010generic}. They  demonstrated that by using time metrics as an objective metric for web applications, the IQX equation yields a QoE mapping function:

\begin{equation}
    \text{QoE}^\text{IQX}(t)=\alpha e^{-\beta t}+\gamma.
    \label{equation:iqx}
\end{equation}
where $t$ is the waiting time metric, $\alpha$, $\beta$ and $\gamma$ are an empirically derived constants. The constants are tuned in accordance with the context (i.e Web, VOIP, Video). 

Egger et al. developed a Web QoE estimation model based on an assumption that the relationship between \textit{W}aiting time and its \textit{Q}oE evaluation on a linear ACR scale is \textit{L}ogarithmic (WQL)~\cite{egger2012time}. This is based on the Weber{-}Fechner law relating a physical stimulus to a perceived change. WQL uses the following expression to estimate Web QoE:
 
 \begin{equation}
\text{QoE}^\text{WQL} = a-b\textrm{ln}(t)
\end{equation}

where $t$ refers to waiting time metric associated with web application,  $a$ and $b$  are computed by minimizing the least square errors between the fitting function and the Mean Opinion Score (MOS) values.
In~\cite{hobetafeld2018speed}, Hossfeld et al. evaluated the accuracy of QoE models using time integral metrics bounded to the ATF time. The authors demonstrated that the time integral metrics can be used in QoE models as a proxy for user perceived waiting times of ATF contents. They observed that estimating ByteIndex using IQX and WQL models yield results with similar accuracy.
%A key challenge  behind  determining ATF time is that if a website has an animated element (\ie video, animation, advertisement), the estimated ATF time and PLT will be equal~\cite{brutlag2011above}. The users, however, may perceive  ATF with a lower VC progress.

% \todo{giving a breif review of web page metrics used to estimate Web QoE}\\

% \todo{Expand more on the visual metrics. }\\

% \todo{Clearly explain what does above the fold mean}\\

% \todo{prepare a figure that shows a time line of loading a page. the different between ATF vs PLT}\\

% \todo{use the paper written by dario tobi and florian and show their finding  that the visual metrics better correlate with the QoE}\\

% \todo{use the  paper written by Dario on the ATF calculation, the plugin for chrome}\\

% \todo{report papers that investigate visual metrics / QoE}\\

% \todo{Explain - here on in the introduction - developers/designers are using animations to handle loading time improve aesthetics etc. heavy javascript and css files, javascript platform makes the websites slower over slower bandwidth  }

% \todo{support the gap in the litterateur using recent papers}\\

%thresholded VC used in a QoE model, it may result in an inaccurate estimation of the QoE. This indicates a need to understand how the web users perceive ATF and to what extent the content and network factor influence the perception of ATF.

\section{Methodology}
\label{methodology}
% (1) to establish a range of annotated visual completion  and above-the-fold (VC ) times. 
%(2) How the network factor impacts the perception of ATF 
%(3) to evaluate the level of  inter-annotator agreement  on the visual completion point. 
%-------------------------------
\subsection{Research Goal and Questions}
\label{sec:method:rq}
The goal of this research is to asses the impact of increased loading time of web page on the perception of ATF for websites with and without animated content. We wish to understand to what extent the users' perception of ATF differs from the systematically computed ATF (objective ATF) using MPHD analysis.
To conduct this study, we formulate the following research questions:

\begin{itemize}[leftmargin=3em]
    \item[\textbf{RQ1:}] Can we annotate perceived ATF based on VC time and progress to establish a range in subjective ATF precision?
    \item[\textbf{RQ2:}] How does network bandwidth impact the visual perception of ATF? 
    \item[\textbf{RQ3:}] Does animated content in the ATF area influence the perception of ATF?
\end{itemize}

\begin{table}[t]
\centering

\begin{tabular}{lp{4.0cm}}
\hline

\multicolumn{1}{c}{\rule{0pt}{1em}%
Questions} &
\multicolumn{1}{c}{Options}  \\ \hline
Utilization of PC &  Basic, Middle, Very good
 \\ \hline
Frequency of Using Websites & Daily, Weekly, 
Less than Monthly \\ \hline

Age & 18-20, 21-29, 30-39, 40-49, 50-59, 
60 or older
 \\ \hline
Gender & Male, Female, Other \\ \hline
Education & High school, Bachelor degree, Graduate degree, Other \\ \hline
Eye Vision & Normal or Corrected Vision, other \\ \hline

\end{tabular}
\caption{Users Characteristics Questionnaire. the questionnaire is prepared using HTML form and filled prior to the user study.}
\label{tab:questionarie}
\end{table}
\begin{table}[t]
\begin{center}

\begin{tabular}{llcl}
\hline
\multirow{2}{*}{Website} & \multicolumn{2}{c}{Content Characteristics}                         \\
\cline{2-3}
                       & Category  & Animated Element \\
\hline
www.aldi.ie                     & Shopping             & \checkmark                  \\
www.independent.ie                   & News             &              \\
www.gumtree.ie                    & Classified             &                  \\
www.theaa.ie                    & Tourism             & \checkmark           \\
www.harveynorman.ie                    & Shopping             & \checkmark            \\
www.autoexpress.co.uk                    & Service             &            \\
www.next.ie                    & Fashion                 &      \\

\hline
\end{tabular}
\renewcommand{\arraystretch}{1}
\end{center}
\caption{Characteristics of  websites}
\label{tab:web}
\vspace*{-4mm}
\end{table}

\subsection{Study Process}
We recruited subjects from a graduate program of an educational institute. We formed a single user study session in a classroom environment. The user study session started with a presentation briefing regarding the ATF concept and definition. It was explained that ATF time refers to the time that the visible part of a web page is populated on the screen. Prior to the experiment the participants received training on how to use the test platform to annotate their perceived ATF time for a web page load video. We demonstrate that the participants will watch a series of videos of loading process of several websites and they have to pause the video on the frame where they believe ATF has occurred. It was emphasised that once the video is paused, it is important to adjust the frame and select the exact frame that you believe best corresponds to the  ATF time, which occurs at a point before the end of all videos. We also informed them of the study consent notice and that participating in this study is optional and that they could withdraw from the study at any point. We then asked the users to navigate to the test platform URL using their laptops and begin the test by filling the questionnaire. The test duration for each participant lasted 15 minutes approximately. Note that in the design of this study, we followed the ITU-T G.1031 recommendation in terms of perceptual dimensions and the system influential factors.

\subsection{Websites and Characteristics}
The complexity of websites and their content features are important aspects of a website quality assessment study. A website can be as simple as a text page (\ie low complexity) or a shopping website with numerous graphical and animated elements (\ie high complexity). The complexity of a web page refers to the number and size of elements of the web page, and is expected to impact the speed of the websites under different network conditions~\cite{schatz2014annotated}. 
As shown in Table~\ref{tab:web}, we have selected seven Irish websites from the top ranked websites listed in SimilarWeb \footnote{\url{https://www.similarweb.com/}}. The selected websites are close to real web browsing experience (\ie websites are familiar to the cohort) and contain content the subjects interact with in their day-to-day life. 
The selected websites are from different content categories: news, shopping, travel, service, classified and fashion. Three websites out of seven have at least one animated HTML element (namely ads, rotating shopping item or a changing banner). The level of complexity of each website differs from each other (\ie number of elements, graphical and animated objects). %\ie each website has a different PLT associated with.  

\begin{figure}[!tp]
\centering
\includegraphics[clip,width=\columnwidth]{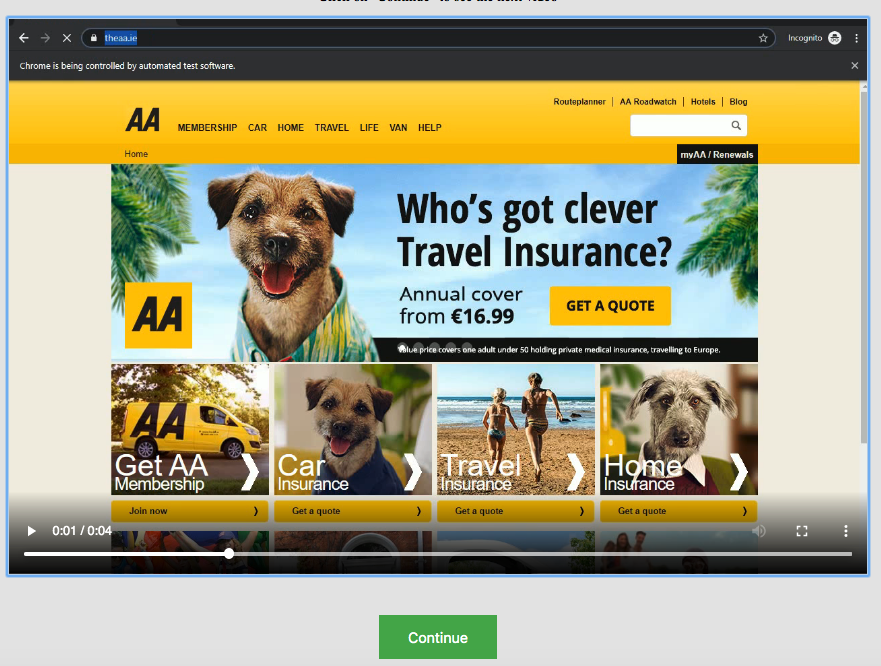}

%\captionsetup{belowskip=0pt}
\caption{Experimental Platform: the users are instructed to pause the video on the frame that they believe ATF is completed. Test subjects use the slider to pinpoint their chosen frame for ATF complete. 
}
\label{fig:plat}
\vspace*{-5mm}
\end{figure}

\subsection{Samples Preparation and Presentation}
To prepare the video of web browsing for the study, we have developed  and used a web recording tool called WebRec\footnote{\url{https://github.com/hzjahromi/webrec}}. The tool is a python script that reads a list of URLs from a CSV file, opens Chrome browser in Incognito's mode and records the video of loading process of each website. The recorded video is 10 frames per second and starts from the time that a URL is submitted until the PLT event.  
We have considered three network bandwidth conditions (1~Mbps, 3~Mbps and 10~Mbps) to influence the loading speed of the websites. The network bandwidth was capped using native Linux traffic control tool ($tc$). For each network condition, we used WebRec to record the video of loading process for all seven websites. In total, we recorded 21 web browsing videos (test cases).  
 In order to present the test cases to the users, we have developed a web-server application \footnote{\url{https://github.com/hzjahromi/atf}}.  The web application presents a questionnaire (see Table \ref{tab:questionarie}) page followed by presentation of a page per pre-recorded video, one-by-one, in a randomised order. For each test case, users have an option to pause the video and select the right frame to label the ATF time (see Fig. \ref{fig:plat}). Finally, the collected data is stored in a CSV file format on the server, with an anonymous but unique user ID.

\section{Experimental Results}
\label{results}
\begin{figure}[!tp]
\centering
\includegraphics[clip,width=0.85\columnwidth]{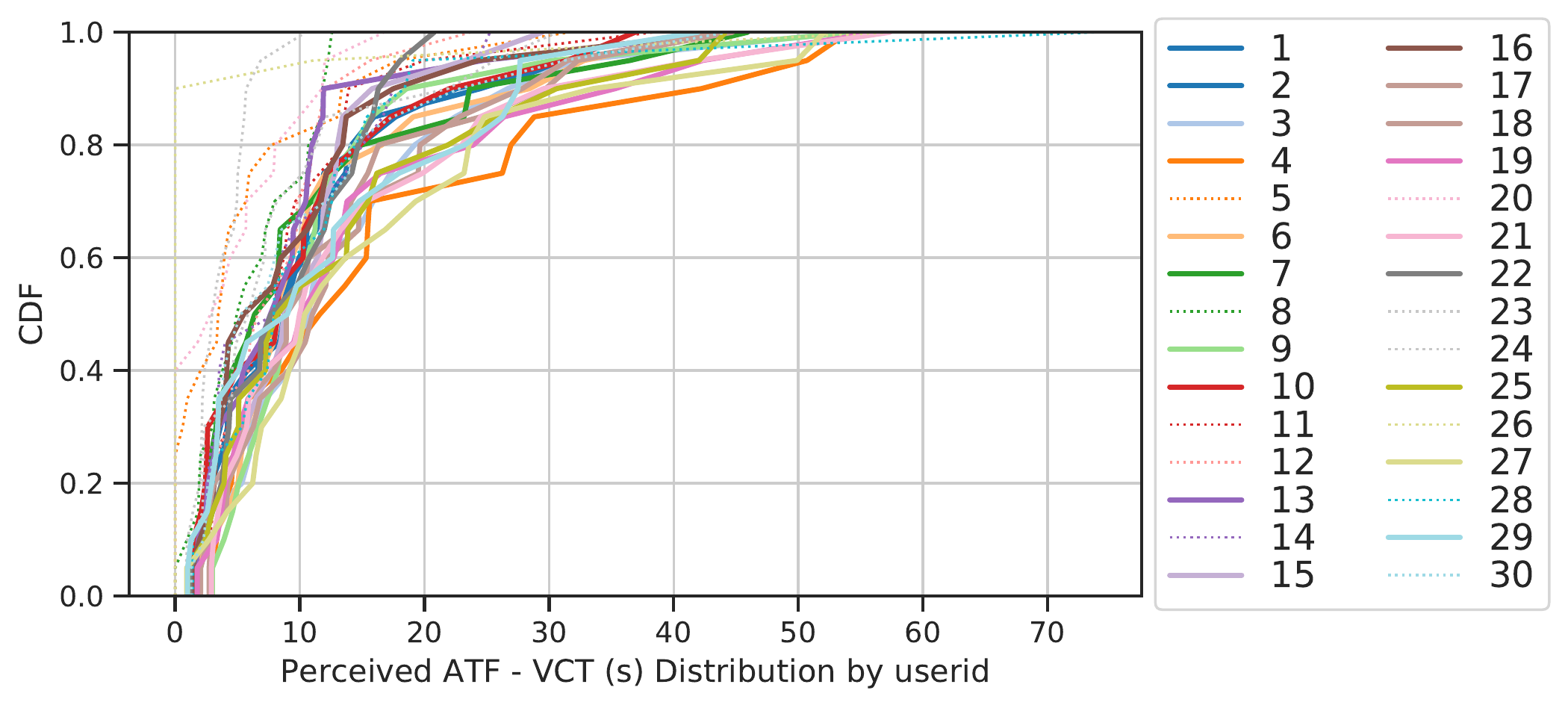}
\includegraphics[clip,width=0.85\columnwidth]{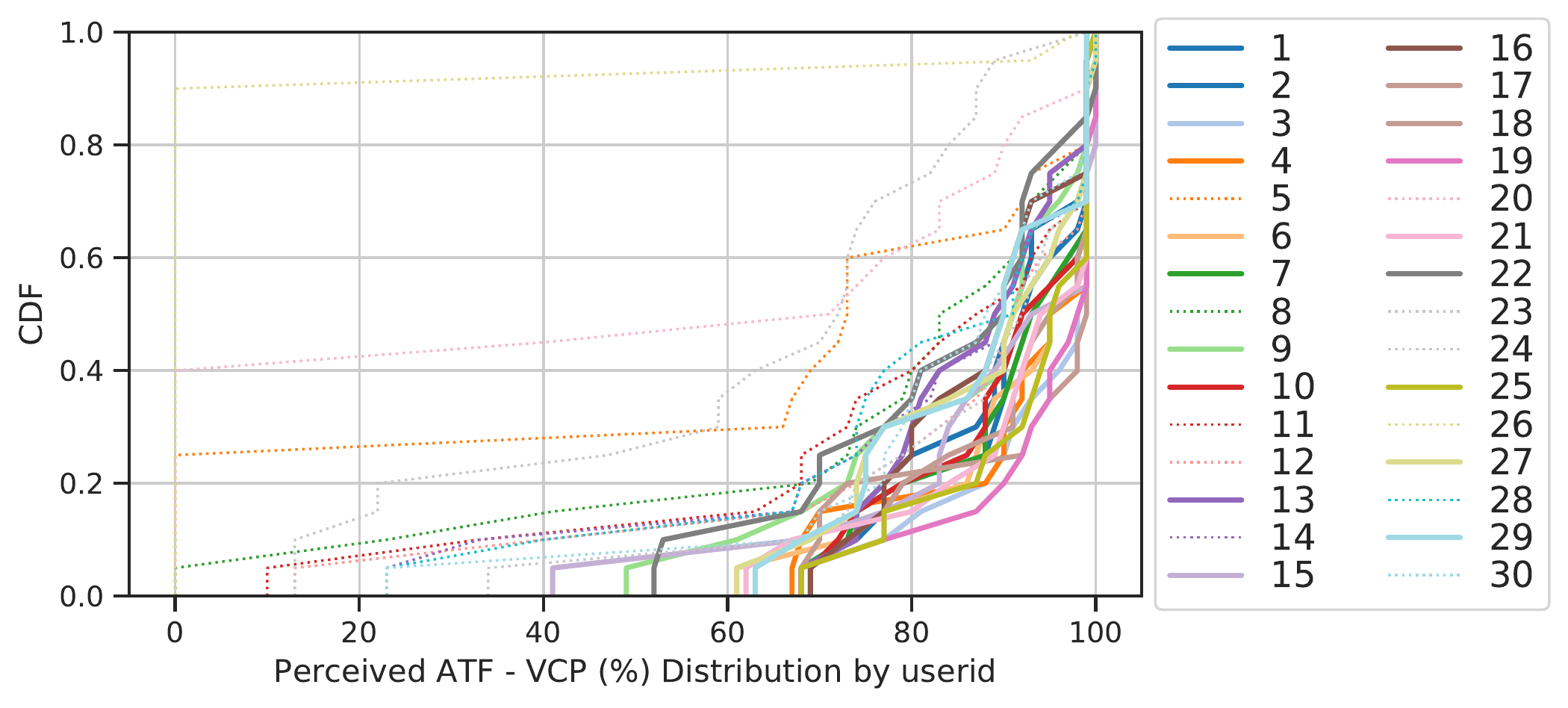}
%\captionsetup{belowskip=0pt}
\caption{The cumulative distribution function (CDF) plot of perceived ATF per ``userid''. The top subplot represents the perceived ATF based on VC time (s). The bottom subplot demonstrates perceived ATF in accordance with VC progress (\%). VC progress is computed based on the users rating of perceived ATF VC time.  The  detected outliers (excluded from analysis) are represented as dotted lines. }
\label{fig:cdf:dist_userid}
\vspace*{-5mm}
\end{figure}

A total of 30 subjects participated in the ATF assessment study. We validated our assumptions with respect to the homogeneity of the test cohort's experience with websites. The cohort consisted of 10 female subjects and 20 male, with an average age-group of 25-35 years old. The majority of participants had higher education backgrounds: 65\% bachelor degree and 35\% declared having a graduate degree. All subjects had normal or corrected eye vision. The survey of subjects explored prior experience with skills in utilising PC and frequency of using websites. 89\% of participants had good level of skills in utilisation of PC and the remaining 11\% reported to have mid-level skills in utilising PC.
Regarding frequency of using websites, 100\% of participants reported to have a daily interaction with websites. 

As part of our data pre-processing step, we detected 11 participants as outliers and excluded them from our subsequent analysis. We generated CDF (cumulative distribution function) plot to demonstrate the distribution of the user's ratings and behaviour (see Fig.~\ref{fig:cdf:dist_userid}). 
The outliers were participants who have not followed the methodology explained in Section~\ref{methodology}, e.g. they chose the beginning or end of the video. They are plotted as dashed lines in Fig.~\ref{fig:cdf:dist_userid}. Furthermore, based on our in-lab pilot study, we determined a minimum VC progress of 40\% is required to have a meaningful paint visualised on the screen. From the excluded participants, 4 subjects had at least one ATF rating with a 0\% VC progress (\ie did not watch or skipped the video) and 7 subjects had a rating with VC progress less than 40\%. It is important to note that, VC progress is computed based on the MPHD analysis of video frames explained in Section~\ref{relatedwork}. 

\begin{figure}[!tp]
\centering
\includegraphics[clip,width=\columnwidth]{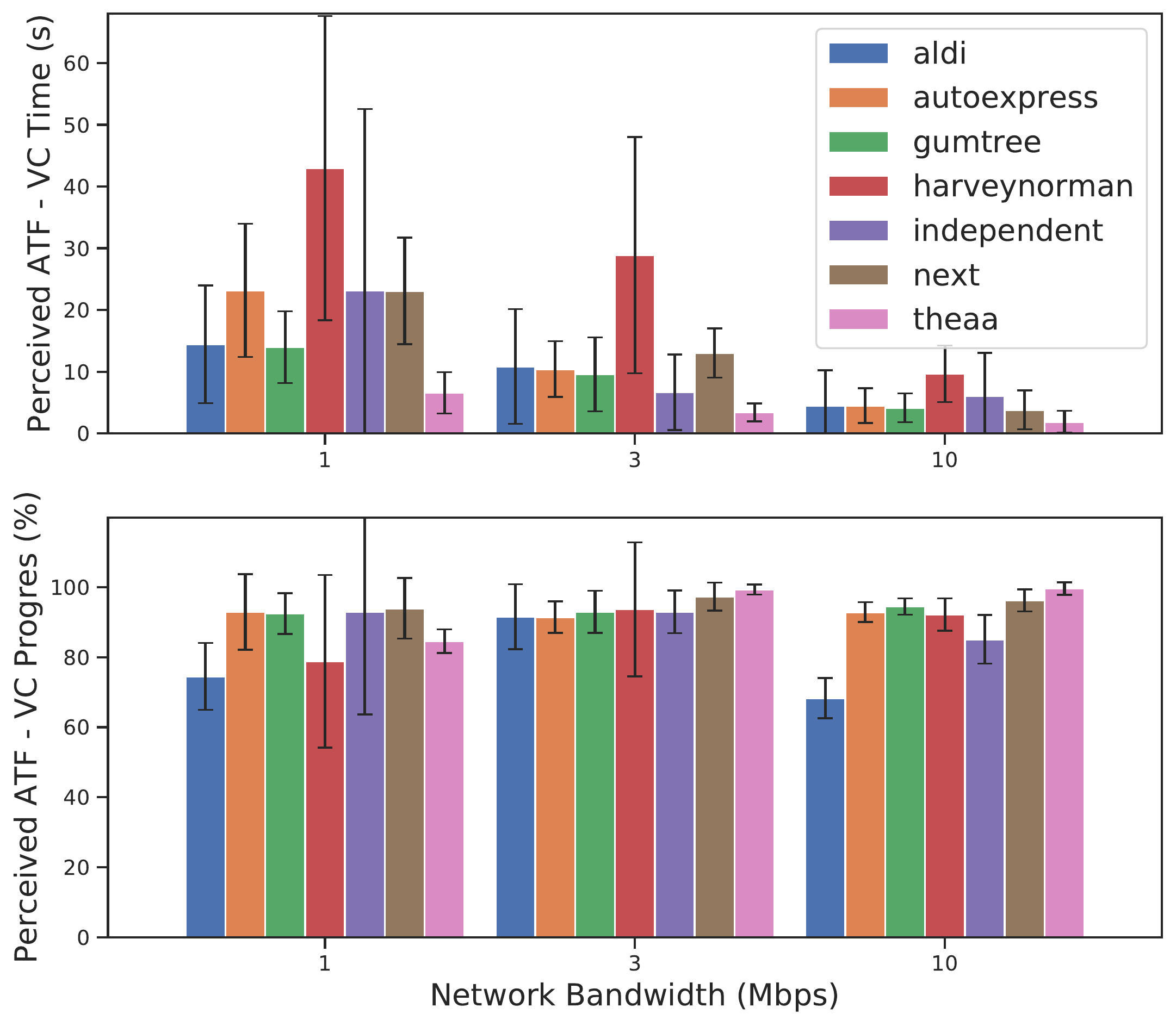}

%\captionsetup{belowskip=0pt}
\caption{User's perceived ATF rating grouped by network bandwidth condition. The top subplot shows the user's rating based on VC time (s). The bottom subplot represents the corresponding VC progress computed based on the perceived ATF VC times.}
\label{fig:raw}
\vspace*{-5mm}
\end{figure}
% The remainder of this section answers the three research question (see section \ref{sec:method:rq}), including underlying analysis.  

\subsection{Establishing a range of annotated Perceived ATF (RQ1)}
We observed that, considering all the websites, the mean value of VC progress corresponding to perceived ATF time is 90.23\% with the standard deviation of 11\% and the median of 93.0\%. Thus, in average, the users perceive ATF  when the website has at least 90\% VC progress. 
 Figure~\ref{fig:raw} shows how the users perceive ATF for seven websites under various loading conditions (caused by network bandwidth), \ie reducing the available network bandwidth, the loading speed of a web page decreases. The top bar plot displays the users' rating of perceived ATF in accordance with a VC time, \ie the point in time during the web page loading process that ATF is perceived. In the bottom subplot, VC time is mapped to a corresponding VC progress. 
If we compare perceived ATF VC time and progress subplots (Fig.~\ref{fig:raw}), it can be seen that, in general, an increase in VC time does not compromise the perception of ATF from the VC progress perspective. The exceptions to this are  explored in the following subsections.  

\begin{figure}[!tp]
\centering
\includegraphics[clip,width=0.97\columnwidth]{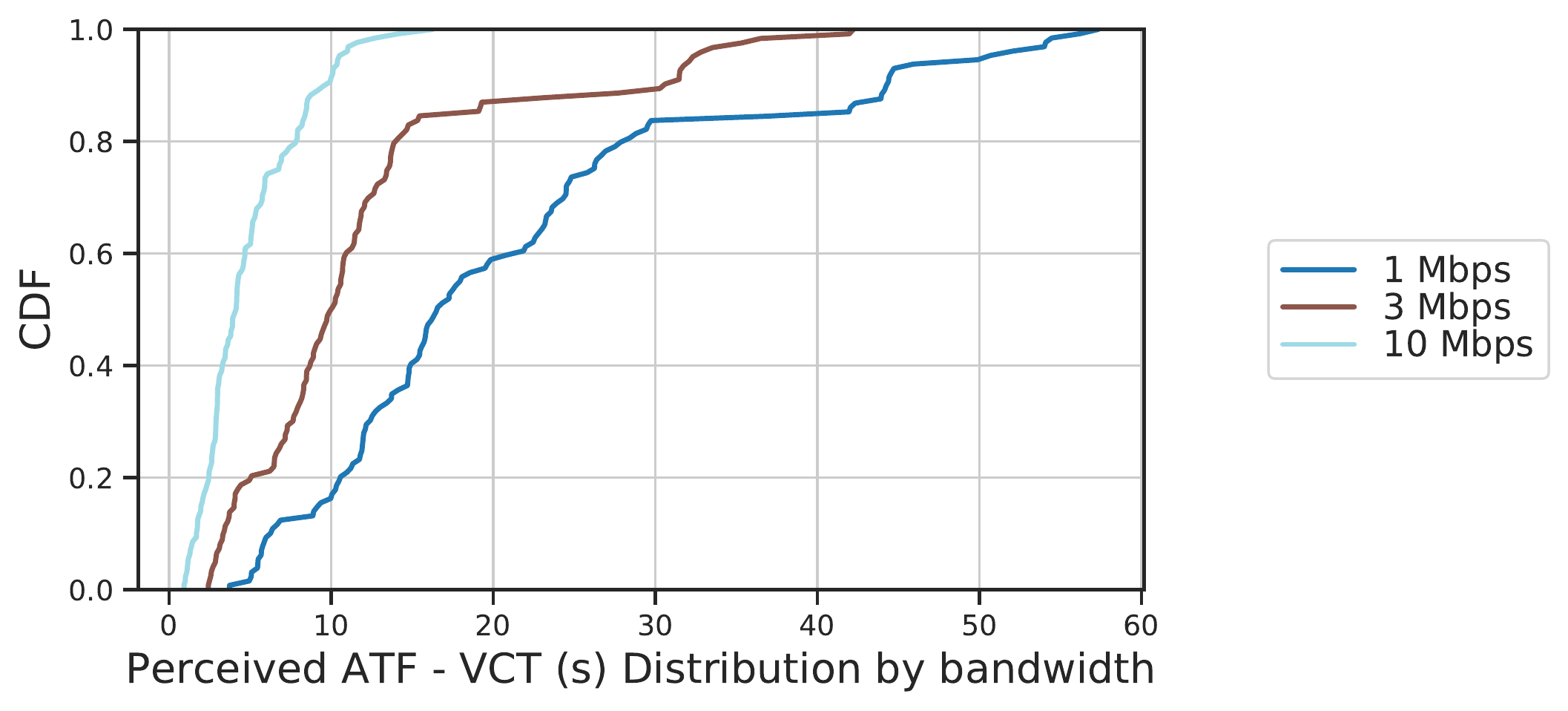}
\includegraphics[clip,width=0.97\columnwidth]{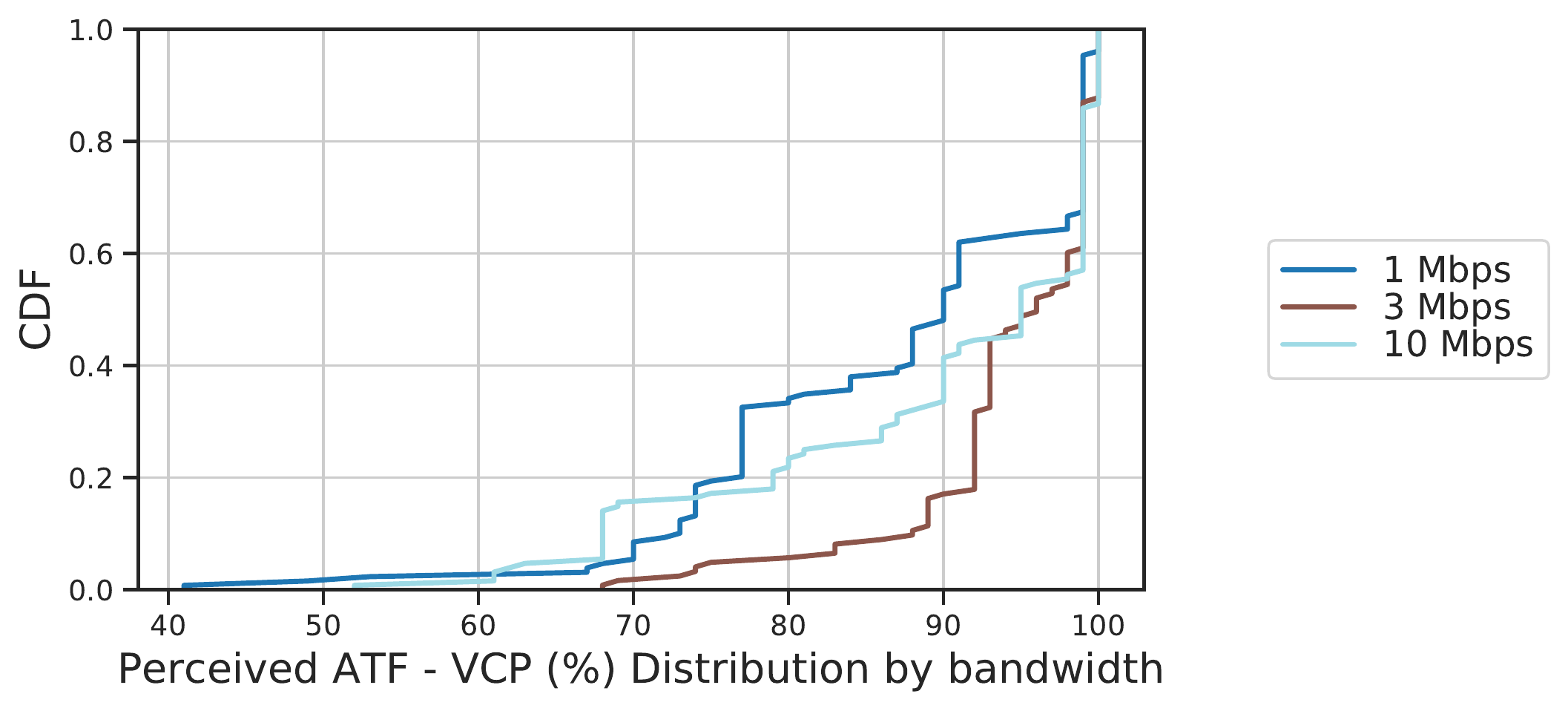}
%\captionsetup{belowskip=0pt}
\caption{The CDF plots of perceived ATF per bandwidth condition  (1,3 and 10 Mbps). The top subplot represents the perceived ATF based on VC time (s). The bottom subplot demonstrates perceived ATF in accordance with VC progress (\%).}
\label{fig:cdf_network}
\vspace*{-5mm}
\end{figure}

\subsection{Influence of network factor on the visual perception of ATF (RQ2)}
An increase in the loading time of ATF content caused by limited network bandwidth does not significantly influence the visual perception of ATF (based on VC progress). However, if we include content factors, the perception of ATF differs depending on the website. 
The top subplot of Fig.~\ref{fig:cdf_network} shows that different network bandwidth levels (10 Mbps, 3 Mbps and 1 Mbps) impact the perceived ATF from the VC time aspect, confirming that reducing network bandwidth increases the perceived ATF time. This is not a surprising finding. However, the overall trend of the bottom subplot of Fig.~\ref{fig:cdf_network}, VC progress, does not show a strong influence of different network bandwidth levels on the visual perception of ATF Our Mixed Model ANOVA analysis (Table~\ref{tab:mm_anova}) also confirms that while the influence of $bandwidth$ on the perceived ATF VC progress is not statistically significant ($p=0.068$), the influence of $website:bandwidth$ on the perceived ATF VC progress is statistically significant ($p<0.001$).   

% latex table generated in R 3.6.2 by xtable 1.8-4 package
% Mon Feb 17 14:51:18 2020
\begin{table}[ht]
\centering
\begin{tabular}{p{4.9cm}rr@{\hskip 0.1cm}l}
  \hline
 Fixed & $F$ & $p$ \\ 
  \hline
website & 18.313 & $<$0.001 & *** \\ 
bandwidth  & 2.890 & 0.068 & \\ 
website:bandwidth & 25.265 & $<$0.001 & *** \\ 
  \hline
 Random & $\chi^2$ &  $p$ \\ 
  \hline
website:userid & 47.504 & $<$0.001 & *** \\ 
bandwidth:userid & 3.402 & 0.065 & \\ 
userid & 0.028 & 0.867 & \\ 
   \hline
\multicolumn{4}{l}{\scriptsize{{*** }$p<0.001$, {** }$p<0.01$, {* }$p<0.05$}}
\end{tabular}
\caption{Mixed-Model ANOVA results for Fixed (F-Test) and Random Effects (Likelihood-Ratio Test). Target variable: Perceived ATF - VC Progress. Asterisks indicate levels of statistical significance. The $F$ is the ratio of the mean-square value and $\chi^2$ refers to the chi-square test.} 
\label{tab:mm_anova}
\end{table}

% latex table generated in R 3.6.2 by xtable 1.8-4 package
% Mon Feb 17 14:51:18 2020
\begin{table*}[ht]
\centering
\begin{tabular}{lrrrrrrrl}
  \hline
 & Estimate & Standard Error & DF & t-value & Lower CI & Upper CI &  p-value.adjust & \\ 
  \hline
  bw 1 - 3 & -2.086 & 0.948 & 36.900 & -2.200 & -4.006 & -0.166 &  0.102 & \\ 
  bw 1 - 10 & -0.240 & 0.941 & 35.900 & -0.250 & -2.148 & 1.668 &  1.000 & \\ 
  bw 3 - 10 & 1.847 & 0.947 & 36.900 & 1.950 & -0.073 & 3.766 &  0.177 & \\ 
  website:bw  aldi 1 -  aldi 3 & 16.997 & 2.054 & 231.100 & 8.280 & 12.951 & 21.043 &  $<$0.001 & *** \\ 
  website:bw  aldi 1 -  aldi 10 & 23.207 & 2.054 & 231.100 & 11.300 & 19.161 & 27.253 & $<$0.001 & *** \\ 
  website:bw  autoexpress 1 -  autoexpress 3 & 1.476 & 2.211 & 244.700 & 0.670 & -2.879 &  0.505 & 1.000 & \\
  website:bw  autoexpress 1 -  autoexpress 10 & 0.124 & 2.092 & 234.200 & 0.060 & -3.998 & 4.245 &  1.000 & \\ 
  website:bw  gumtree 1 -  gumtree 3 & -0.662 & 2.054 & 231.100 & -0.320 & -4.708 & 3.385 &  1.000 & \\ 
  website:bw  gumtree 1 -  gumtree 10 & -2.215 & 2.054 & 231.100 & -1.080 & -6.261 & 1.831 &  1.000 & \\ 
  website:bw  harveynorman 1 -  harveynorman 3 & -14.842 & 2.019 & 228.100 & -7.350 & -18.820 & -10.864 &  $<$0.001 & *** \\ 
  website:bw  harveynorman 1 -  harveynorman 10 & -13.129 & 2.092 & 234.100 & -6.280 & -17.250 & -9.008 &  $<$0.001 & *** \\ 
  website:bw  independent 1 -  independent 3 & 0.576 & 2.128 & 237.500 & 0.270 & -3.616 & 4.768 &  1.000 & \\ 
  website:bw  independent 1 -  independent 10 & 7.820 & 2.054 & 231.100 & 3.810 & 3.774 & 11.867 &  0.042 & *\\ 
  website:bw  next 1 -  next 3 & -3.490 & 2.054 & 231.100 & -1.700 & -7.536 & 0.556 & 1.000 & \\ 
  website:bw  next 1 -  next 10 & -2.481 & 2.092 & 234.100 & -1.190 & -6.603 & 1.640 & 1.000 & \\ 
  website:bw  theaa 1 -  theaa 3 & -14.659 & 2.128 & 237.500 & -6.890 & -18.851 & -10.467 &  $<$0.001 & ***\\ 
  website:bw  theaa 1 -  theaa 10 & -15.005 & 2.054 & 231.100 & -7.310 & -19.051 & -10.959 &  $<$0.001 & ***\\ 
  website:bw  aldi 3 -  aldi 10 & 6.210 & 2.019 & 228.100 & 3.080 & 2.233 & 10.188 &  0.504 \\ 
  website:bw  autoexpress 3 -  autoexpress 10 & -1.352 & 2.134 & 237.400 & -0.630 & -5.556 & 2.853  & 1.000 & \\ 
  website:bw  gumtree 3 -  gumtree 10 & -1.553 & 2.089 & 234.200 & -0.740 & -5.669 & 2.562 &  1.000 & \\ 
  website:bw  harveynorman 3 -  harveynorman 10 & 1.713 & 2.092 & 234.100 & 0.820 & -2.408 & 5.835 &  1.000 & \\ 
  website:bw  independent 3 -  independent 10 & 7.245 & 2.092 & 234.100 & 3.460 & 3.123 & 11.366 &  0.126 & \\ 
  website:bw  next 3 -  next 10 & 1.009 & 2.128 & 237.500 & 0.470 & -3.183 & 5.201 & 1.000 & \\ 
  website:bw  theaa 3 -  theaa 10 & -0.346 & 2.092 & 234.100 & -0.170 & -4.468 & 3.775 & 1.000 & \\ 
  \hline
 \multicolumn{5}{l}{\scriptsize{{*** }$p<0.001$, {** }$p<0.01$, {* }$p<0.05$}}
\end{tabular}
\caption{ The conjoint analysis based on differences of Least-Square Means ~\cite{bak2009conjoint} for the Perceived ATF VC progress.  Asterisks indicate levels of statistical significance. \textit{p-value adjusted} represents a significant differences between factor levels using ``bonferroni-corrected'' method. \textit{CI} refers to the confidence interval and \textit{DF} refers to the degree of freedom.
} 
\label{tab:lsmeans}
\end{table*}

\subsection{The influence of animated content on the perception of ATF (RQ3)}
Based on our post-hoc analysis (see Table~\ref{tab:lsmeans}), we have observed that an increase in ATF load time caused by network bandwidth, impacts the visual perception of ATF (VC progress) for websites with animated contents (marked in Table~\ref{tab:web}).  
Table~\ref{tab:lsmeans} shows that for the three websites with animated content, the influence of increased ATF time (cause by different bandwidth levels) on the visual perception of ATF (VC progress) is statistically significant ($\text{p-value.adjusted}<0.001$). For instance, from Fig.~\ref{fig:raw}'s lower subplot, it can be seen that the value of perceived ATF VC progress for \textit{www.harveynorman.ie} decreases when the perceived ATF VC time increases. The same is not true for \textit{www.gumtree.ie} which does not have animated content. 
It is interesting to note that, for the websites with animated elements, the perceived ATF VC progress does not always decrease when the ATF VC time increases. Unlike \textit{www.harveynorman.ie} and \textit{www.theaa.ie} that show a decreasing trend, the user ratings show an increasing trend for \textit{www.aldi.ie}.

\section{Discussion}

\label{sec:discussion}
In the previous sections we have explored how web application users perceive ATF for two aspects: time and progress with pages including animated and static content. In Fig.~\ref{fig:raw} we have quantified the magnitude of change in the perceived ATF with regards to the VC time and progress. The visual perception of ATF for the web sites without animated content is nearly consistent for the wide range of cases tested (ATF is perceived at approximately 90\% of VC progress). However, for the websites with animation we see a different trend among the websites. For instance, \textit{wwww.harveynorman.ie} and  \textit{wwww.theaa.ie} show a downward trend for perceived ATF as bandwidth increases. Surprisingly, \textit{wwww.aldi.ie} shows the opposite trend. We think that the mixed trend in the perception of ATF can be further explained by the content and human psychological factors as we will now discuss. %Regarding the content factor, the difference in the mean pixel histogram of the animated frames can influence the VC progress estimation~\cite{gao2017perceived}.
Human perception of waiting time is also  fluid and can get manipulated by visual effects~\cite{harrison2010faster}. Web users can start losing attention after two seconds of waiting time without any visual progress~\cite{szameitat2009behavioral}. However, users are willing to wait more if the website has an animated content with a smooth visual progress (e.g. using ``lazy loading'' technique~\cite{park2019design}). Similarly, in~\cite{kuisma2010effects}, the authors confirm that the content of animation has a positive effect on the attention of the web users, e.g. for an animated picture of a skyline containing ``skyscrapers'' it has a positive effect on attention to the buildings, but for banners and advertisements animation has a negative effect on attention. It can thus be suggested that the content of animation, the speed of animation and the loading progress can all influence the perception of ATF.  The question arises as to what extent the perception of ATF for animated and non-animated contents is impacting the applications of ATF metrics? In the following subsections we explore the accuracy of objective ATF, SI and QoE estimation models with reference to the perceived ATF collected in this user study. The term ``objective ATF'' is used to refer to the ATF time estimated using MPHD analysis of video frames and associated with 100\% of VC progress (see Fig. \ref{fig:def:siatf}). 
%to utilise correlation plots and  coefficient of determination analysis (known as ''R-squared'') 

\begin{figure}[!tp]
\centering

\includegraphics[width=1\columnwidth]{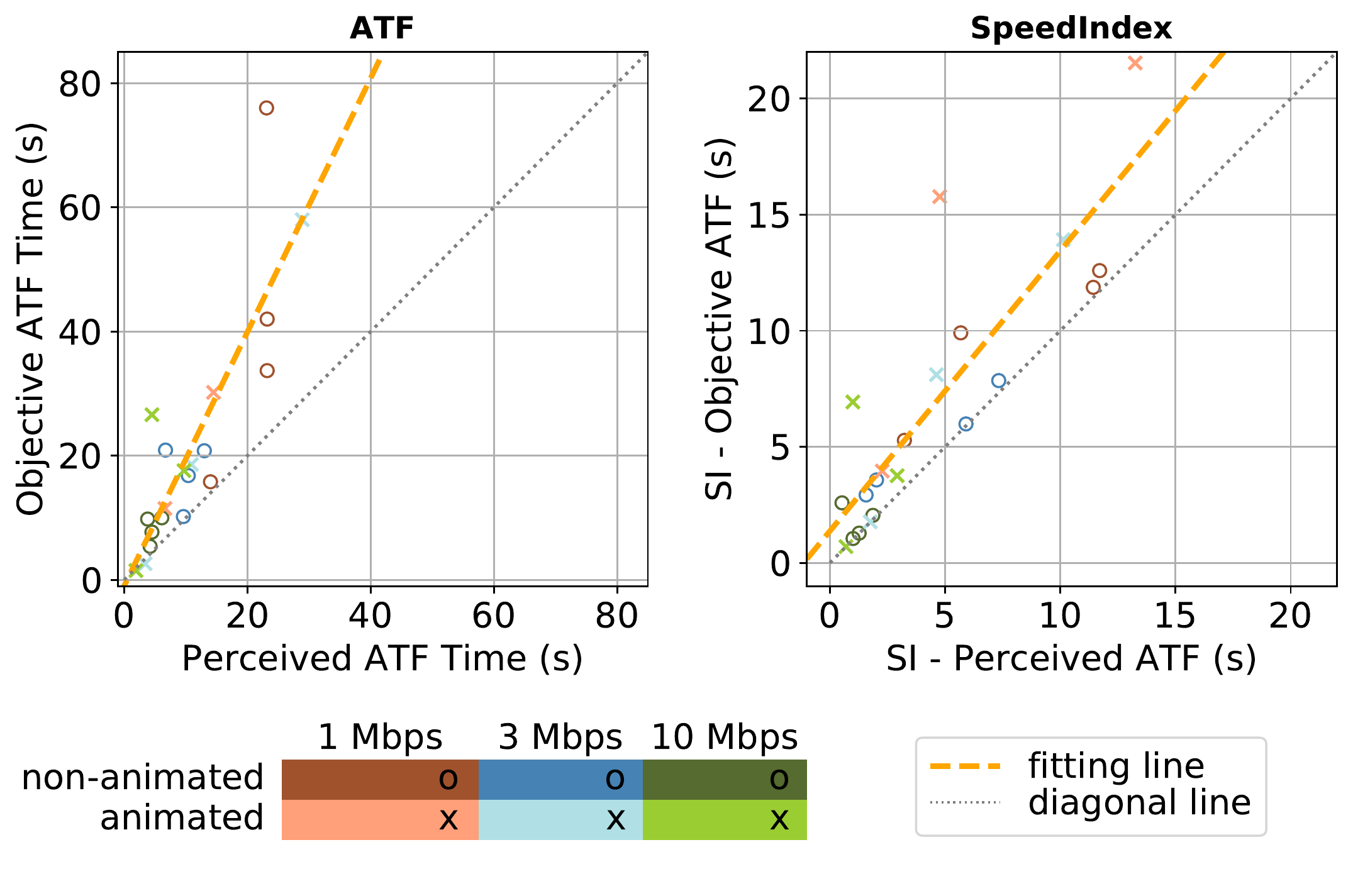}
%\captionsetup{belowskip=0pt}
\caption{The left subplot represents the correlation between perceived ATF time (x-axis) and objective ATF time (y-axis).  The right subplot shows the correlation between SI based on Perceived ATF Time (x-axis) and  SI based on Objective ATF Time (y-axis). The diagonal lines show how identical the output of metrics are.}
\label{fig:siaft}

\vspace*{-5mm}
\end{figure}
\begin{figure}[!tp]
\centering
\centering
\includegraphics[width=1\columnwidth]{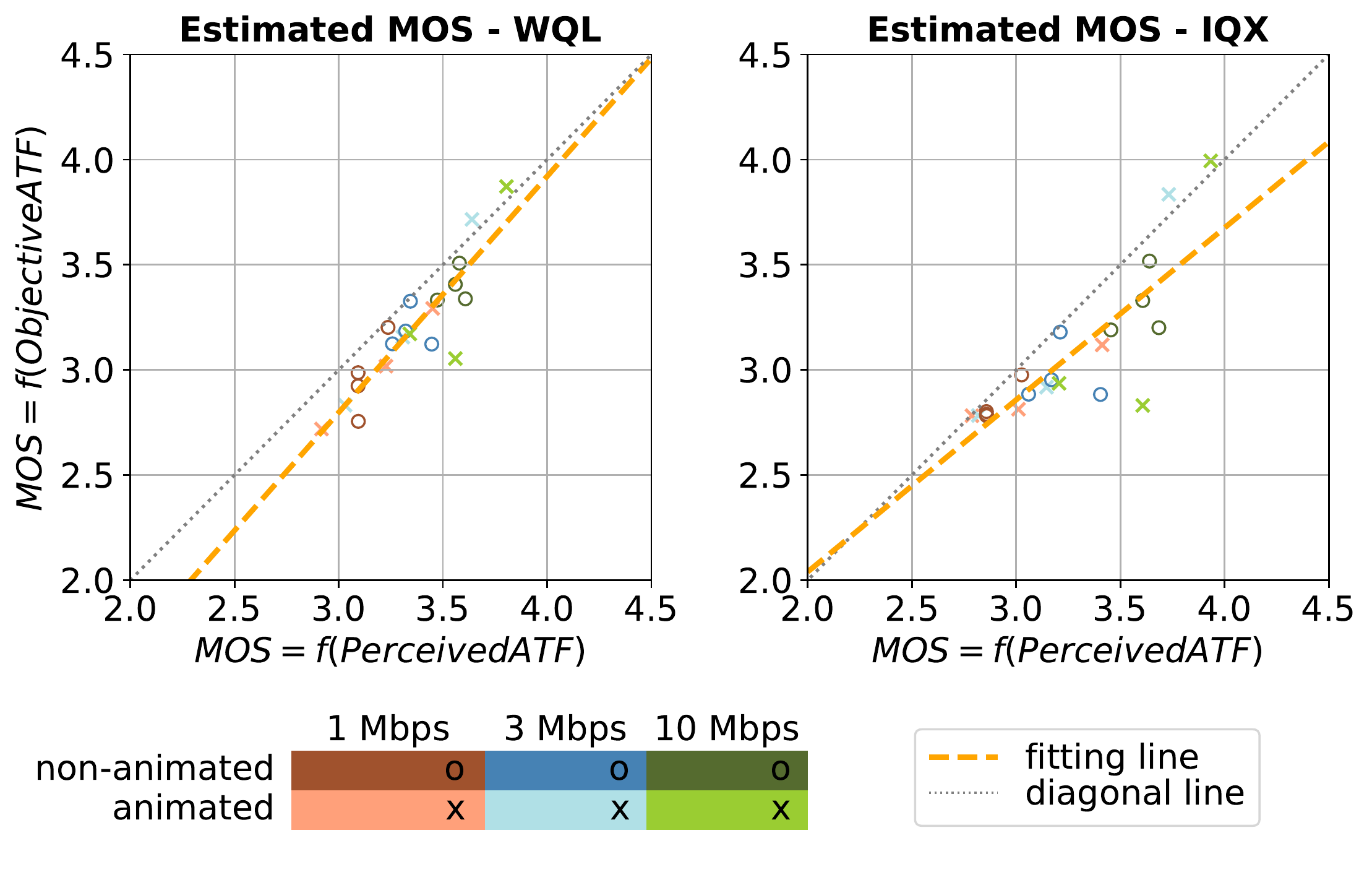}

%\captionsetup{belowskip=0pt}
\caption{Correlation between MOS estimation of perceived ATF (x-axis) and objective ATF (y-axis).  The MOS is approximated using WQL (left subplot) and IQX (right subplot) models.}
\label{fig:qoemodels}

\vspace*{-5mm}
\end{figure}
\subsection{The Approximation of  Objective ATF}
 From the left subplot of Fig.~\ref{fig:siaft}, we can see that in most cases, the objective ATF is over estimating the ATF time, \ie The perceived ATF time is smaller than estimated ATF time. The same holds true for both animated and non-animated contents.  We have used coefficient of determination analysis (known as ''R-squared'')  to quantify how close the perceived ATF time and the objective ATF time are to the fitted regression line. The computed R-squared based on all  test cases is 0.83, meaning that, 83\%  of the variation in perceived ATF can be explained by the objective ATF. From Fig.~\ref{fig:siaft} (left subplot)  we can also see that the distance between two variables increases as the network bandwidth decreases. By considering the magnitude of change between two variables, we understand that the influence of content on the objective ATF estimation is more apparent when the network bandwidth decreases (increased ATF time). 

\subsection{Influence of ATF time on SpeedIndex}
For the majority of the non-animated test cases, the computed SI using perceived and objective ATF yield a similar result (e.g. the proportion of points on the diagonal line of Fig.~\ref{fig:siaft}). However, for the websites with animated content, the SI computed using the objective ATF prediction overestimates the speed of loading (see the wider range of \textsf{x} markers in the right subplot of Fig.~\ref{fig:siaft}). While the overestimation of  speed of loading  for the websites with non-animated content is more apparent for the lower bandwidth levels (3~Mbps and 1~Mbps), for the websites with animated content,  the overestimation  occurs for all bandwidths tested. 
By including all test cases (animated and non-animated), the computed  R-squared value is 0.74. If we separate the  content types and re-compute the R-squared, the computed value for animated and  non-animated will be 0.78 and 0.90, respectively. The lower R-squared value for the animated websites confirms the known issue with SI and VC computation (SI  may over estimate the loading speed of a website, when animated contents exist in the ATF area~\cite{speedindex}). As explained in Section~\ref{relatedwork}, thresholding techniques are commonly used in the literature to improve the accuracy of SI for both animated and non-animated contents, e.g. SI is upper-bounded to the time associated with 90\% of VC progress. %However, thresholding may impact the estimation of QoE models. 

\subsection{ATF and QoE Models}
To understand the robustness of QoE estimation models and how the objective ATF impacts QoE models, we have used constants of IQX and WQL models fitting curves from da Hora et al.~\cite{da2018narrowing} study and, estimated the corresponding MOS value for a given ATF time. We have performed our analysis based on an assumption that the result of da Hora et al. study is generalisable and can be used to estimate QoE for our test cases. Our investigation shows that both WQL and IQX are generally underestimating the perceived quality using the objective ATF (Fig.~\ref{fig:qoemodels}). In both cases, the estimated MOS using objective ATF is lower than the estimated MOS using perceived ATF. However, the MOS estimation using WQL (logarithmic model) has a lower variation than using IQX (exponential model). While WQL  has a consistent behaviour across different bandwidth conditions, the underestimation of IQX increases with higher network bandwidth levels (lower ATF time). \ie sparse data-points in  the right subplot of Fig.~\ref{fig:qoemodels} represents a higher variation in MOS estimation for IQX model. Due to the nature of logarithmic and exponential curve fitting of WQL and IQX,  there is no major effect on the MOS estimation using ATF for the websites with animated and non-animated contents. The computed R-squared for WQL and IQX are 0.79 and 0.66, respectively. The next logical step is to understand how the accuracy of QoE models changes based on SI bounded to perceived and objective ATF. We considered to do the next step, however, in reviewing the literature, we could not find a publicly available dataset that uses SI for QoE estimation. Therefore, it is important to bear in mind that the robustness of QoE models may differ if SI is used as a proxy metric. 
 
  %Different techniques can be used to optimise ATF measurement for the animated contents and improve the accuracy of visual metric. For instance, the animated content can be freezed while computing the ATF time. 

\section{Conclusion and Future Work}
\label{conclusion}
In this paper, we explored the influence of network bandwidth on the perception of ATF for websites with and without animated content. In our experiment we used popular commercial websites homepages, representative of different content categories. Our results show that the perception of ATF does not significantly change for websites without animations. However, for websites with animations the perception of ATF changes as the load time increases (as a result of decreasing the network bandwidth). Content is a key influential factor in  QoE studies. The amount of animation and multimedia content used in websites is increasing. This growth increases the challenges involved in accurate ATF estimation.  Despite the fact that the heuristic  methods work, we postulate that  what the user perceives on the screen may  deviate from what the heuristics tell us.  \ie the heuristics methods does not consider the visual content of the web elements. %We believe that determining ATF time using video frames is  beneficial and relevant to QoE.  Thus, concerning the future work we foresee including the wider range of websites with more content features. 
To this end, we  plan to investigate how the accuracy of ATF  estimation can be improved for the websites with multimedia and animated content, and to improve the robustness of metrics like SI that rely on an accurate ATF estimation to predict load time.
\section{Acknowledgement}
The authors would like to thank Raimund Schatz who supported this study by providing analytical tools, methods and helpful suggestions. This publication has emanated from research supported in part by a research grant from Science Foundation Ireland (SFI) and is co-funded under the European Regional Development Fund under Grant Number 13/RC/2077 and Grant Number SFI/12/RC/2289\_P2.

\Urlmuskip=0mu plus 1mu\relax
\bibliographystyle{./bibliography/IEEEtran}
\bibliography{./bibliography/IEEEabrv,main.bib}
\vspace{12pt}

\end{document}